\documentclass{elsart}

\usepackage{amsmath}        
\usepackage[english]{babel} 
\usepackage{graphicx}       
\usepackage{natbib}

\begin{document}
\begin{frontmatter}

\title{Estimating the angular resolution of tracks in neutrino
telescopes based on a likelihood analysis}
\author{Till Neunh{\"o}ffer}
\ead{Till.Neunhoeffer@uni-mainz.de}
\address{Institute of Physics, University of Mainz, Staudinger Weg 7, 55099 Mainz, Germany}

\begin{abstract}
A semianalytic method to estimate the angular resolution of tracks,
that have been reconstructed by a likelihood approach, is presented.
The optimal choice of coordinate systems and resolution parameters, as
well as tests of the method are discussed based on an application for a
neutrino telescope.
\end{abstract}

\end{frontmatter}

\section{Introduction}

This paper describes a statistical procedure to extract
resolution estimates on a track by track basis. The method
was developed for the AMANDA Neutrino Telescope at the South
Pole~\cite{tillpub:Andres:2001ty}, which uses a 3-dimensional grid of
photosensors imbedded in highly transparent ice to provide spatial and
time resolution of Cherenkov photons, that e.g. arise from long muon
tracks.

The knowledge of track resolutions is of particular importance in the
search for localized sources, such as distant galaxies. The resolution
information can in addition be used to suppress mis-reconstructed tracks,
that typically are less well defined. The method is not limited to muon
reconstruction in neutrino telescopes and can be applied to any
experiment in which tracks have been reconstructed with a likelihood
approach.

\section{Technical aspects of obtaining the confidence ellipse}
\label{section:technicals}

\subsection{Definitions and prerequisites}

In all following paragraphs a {\em pattern of hits\/} in an event will be
modelled by an infinitely long oriented straight
line. It will be parameterized by an arbitrary point $\mathbf{r}$ on the
track and the zenith $\vartheta \in [0,\pi]$ and azimuth $\varphi \in
[0,2\pi]$ angles in a given detector coordinate system.

The procedure presented operates on likelihood functions,
thus a likelihood based reconstruction must be available. For
any given hypothesis $\{\mathbf{r}_0, \vartheta_0, \varphi_0\}$
and an observed pattern of hits $\mathbf{P}$, a likelihood $\mathcal{L}(\mathbf{r}_0,
\vartheta_0, \varphi_0;\mathbf{P})$ must be calculable. The
hypothesis $\mathrm{h}_{\mathrm{best}} = \{\mathbf{r}_{\mathrm{best}},
\vartheta_{\mathrm{best}}, \varphi_{\mathrm{best}}\}$ with the highest value
of $\mathcal{L}=\mathcal{L}_{\mathrm{best}}$ is considered to be the best
estimate for the true values. It is assumed that this extremum has
already been found.

The uncertainties, which are to be determined, are in essence\footnote{A
coordinate transformation needs to be done first, which will be
explained in the following section.} the errors of $\vartheta$ and $\varphi$.
It is mandatory to use the optimal coordinate system for each event and
to consider the errors in the determination of $\mathbf{r}$ as well.

In order to arrive at a confidence ellipse in the two dimensions of track
direction, the set in parameter space is being searched for, where the value of the
negative logarithmic likelihood function has changed by $1/2$ with respect
to $-\log\mathcal{L}_{\mathrm{best}}$:
\begin{equation}
\Delta (-\log\mathcal{L}) = (-\log\mathcal{L}_{\mathrm{ellipse}}) -
(-\log\mathcal{L}_{\mathrm{best}})
\stackrel{!}{=}\frac{1}{2} \qquad .
\end{equation}
After discussing the coordinate systems used, the dimension of the
problem will be reduced to the two dimensions of direction by
eliminating the three dimensions of the location of $\mathbf{r}$.
This is done using a numerical minimization process. The remaining
$\chi^2$-minimization is done analytically to obtain the uncertainty
estimators.

The shape of the negative log-likelihood function around its minimum is
considered Gaussian. The tests described in section~\ref{section:tests}
prove that this assumption is realistic.

\subsection{On coordinate systems}

\begin{figure}[tbp]
\begin{center}
\includegraphics[width=\textwidth]{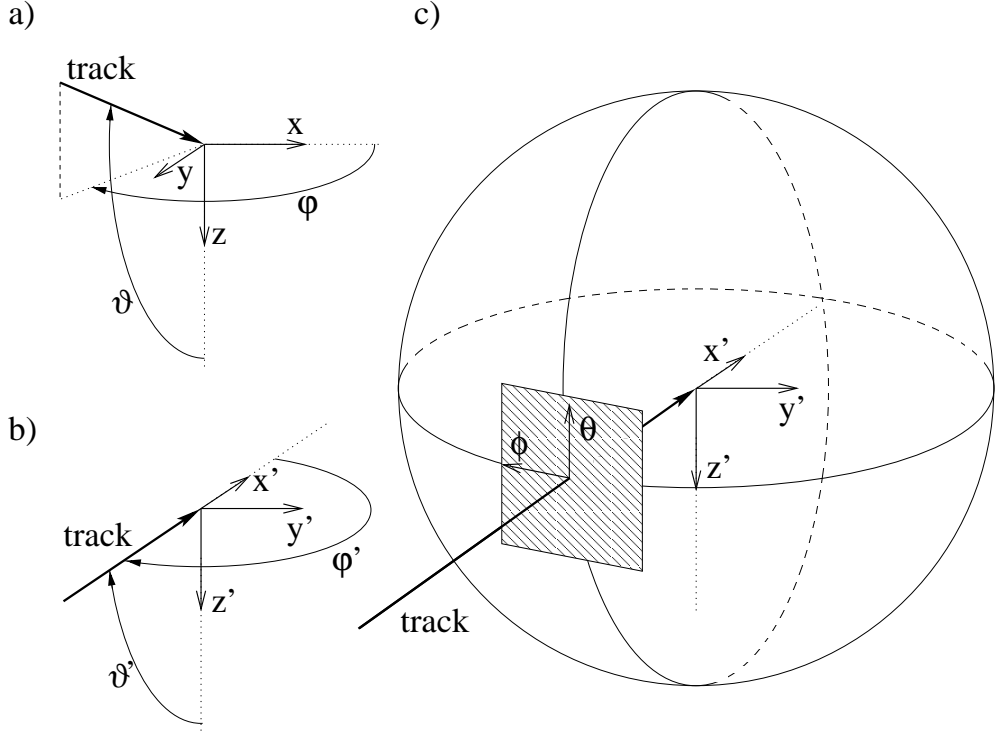}
\caption{Figure a) shows a track in a Cartesian coordinate system $(x,y,z)$.
The direction towards the track's origin is described with standard spherical
coordinates $(\vartheta,\varphi)$. The downward orientation of the $z$-axis follows the AMANDA
standards~\cite{art:recopaper:2004}. In b) a rotated coordinate system
$(x',y',z')$ is
presented, where the $x'$-axis is defined by the track. To describe
directions that are close to the track's direction, relative coordinates
$\theta = \vartheta' - \pi/2$ and $\phi = \varphi' - \pi$ are introduced.
They are approximately the coordinates in the tangent plane, as shown in
c).}
\label{fig:coordinates}
\end{center}
\end{figure}
There are two sets of coordinate systems used in the method presented
here. The first system uses all five coordinates
$(x,y,z,\vartheta,\varphi)$, which are connected to the
3-dimensional Cartesian detector coordinate system as illustrated in
Figure~\ref{fig:coordinates} a). In the second system the problem is
reduced to the 2-dimensional subset of angles $\phi$ and $\theta$.

The complication with $\vartheta$ and $\varphi$ is that the first has
boundaries and the second is periodic. Moreover, at the boundaries
of $\vartheta$ the variable $\varphi$ becomes meaningless. In a
mathematical sense, $\vartheta$ and $\varphi$ are locally orthogonal.
However, if one looks at these coordinates in the area surrounding a point,
they are distorted - in particular close to $\vartheta = 0$ and $\vartheta
= \pi$.

As the behaviour of the likelihood function around the well known
best track estimate $\mathrm{h}_{\mathrm{best}}$ is to be investigated,
the coordinates $\vartheta$ and $\varphi$ should be as close to
being orthogonal as possible {\em at that particular direction\/}
$(\vartheta_{\mathrm{best}},\varphi_{\mathrm{best}})$. Hence the
3-dimensional detector coordinate system is rotated such that
the new x-axis, $x'$, is defined by the track. This results in
$\vartheta_{\mathrm{best}}' = \frac{\pi}{2}$ and $\varphi_{\mathrm{best}}' =
\pi$, as illustrated in Figure~\ref{fig:coordinates} b), where the
distortion is minimal.

Here and in the following section~\ref{section:spaceangle} only
{\em relative\/} coordinates $\theta = \vartheta' - \pi$ and $\phi=
\varphi' - \pi/2$ with respect to $\vartheta_{\mathrm{best}}'$ and
$\varphi_{\mathrm{best}}'$ will be referred to. This can be thought of
as local coordinates in the tangential plane to the unit sphere at
$(\vartheta_{\mathrm{best}},\varphi_{\mathrm{best}})$ with the origin at
the point, where the plane touches the sphere\footnote{This is an
approximation that implies that the coordinates on the surface of
a sphere can be assumed Cartesian for small distances. The maximum
error introduced is of the order of $0.15\%$ for distances less than
$5^\circ$.}. By choosing a proper rotation, it is assured that the
$\theta$-direction is parallel to the $\vartheta$-coordinate, i.e.
towards the negative $z/z'$-axis. The rotation must be such that the
track, the $z$-axis, and the $z'$-axis are in the same plane.

The local coordinates are displayed in Figure~\ref{fig:coordinates} c). 

\subsection{Reducing the dimensionality of the problem}

The location of the point $\mathbf{r}$ is of no concern for the discussion
of where a track points to. Nevertheless the correlations of the errors in
direction with the errors in determining $\mathbf{r}$ need to be considered. 

The reduction procedure of the dimension
shall first be discussed assuming only one spatial dimension $x$ and one
directional dimension $\theta$.

%
\begin{figure}[tbp]
\begin{center}
\includegraphics[width=8cm]{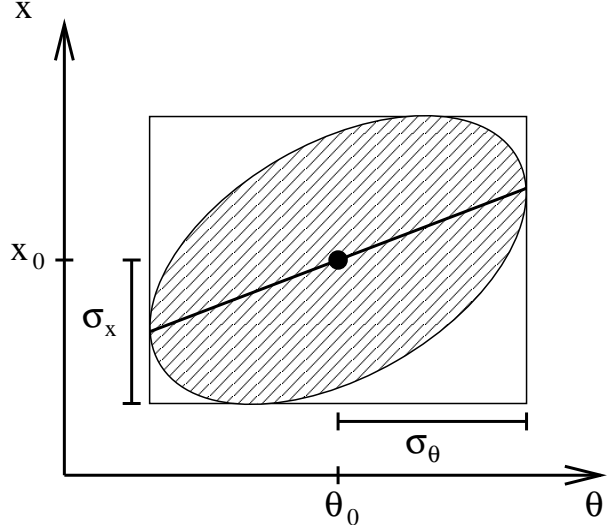}
\caption{
Confidence ellipse in two dimensions with
a directional coordinate $\theta$ and a spatial coordinate $x$. The
rectangle around the ellipse has the size $2\,\sigma_\theta$ by
$2\,\sigma_x$. Finding for each $\theta$ the corresponding $x$ with extremal
likelihood, one arrives at the line shown.}
\label{fig:ellipsewinkel} 
\end{center}
\end{figure}
Consider the two-dimensional confidence ellipse in $(\theta,x)$-space
(Figure~\ref{fig:ellipsewinkel}). It is tilted versus the $\theta$-axis,
which indicates a correlation of $\theta$ and $x$. For each $\theta$ one
determines the value of $x$ for which the common likelihood function
gets maximal. In the strictly Gaussian case, the result is a straight
line. The likelihood function along this line is again a Gaussian as a
function of $\theta$. The width of this Gaussian is $\sigma_{\theta}$
as can be seen by straight forward calculation. In this error estimator
$\sigma_{\theta}$ the correlation $\mathrm{cov}(\theta,x)$ is included by
construction.

The extension to three spatial and two directional dimensions is straight
forward: first one determines a
set of points in $(\phi_i,\theta_i)$. For each point, the best guess, i.e.
the best likelihood $l_i = -\log\mathcal{L}_i$, is found with respect to
$\mathbf{r}$.

In the Gaussian approximation these points $l_i(\phi_i,\theta_i)$ form a
paraboloid. Its parameters are found by analytic minimization, which is
discussed next.

\subsection{The analytic minimization}

In order to find the paraboloid that best fits the points
$l_i(\phi_i,\theta_i)$, an analytic $\chi^2$-minimization is performed. The
paraboloid $f(\phi,\theta;\delta,\beta_1,\beta_2,\Gamma)$ is parameterized in
its polynomial representation:
\begin{equation}
f(\phi,\theta;\delta,\beta_1,\beta_2,\Gamma) = \delta 
+ \beta_1\phi+\beta_2\theta
+  \frac{1}{2} (\phi,\theta)\Gamma
(\phi,\theta)^T 
\qquad.
\end{equation}
The constant parameter is represented by $\delta$ and the parameters
in first order by $\beta_1$ and $\beta_2$. The symmetric $2\times
2$-matrix $\Gamma = \left(\begin{array}{cc} \Gamma_{11} & \Gamma_{12}
\\\Gamma_{12} & \Gamma_{22}\end{array}\right)$ contains the second order
parameters.
Using the method of least squares one calculates
\begin{equation}
  \label{eq:chisquare}
  \chi^2 = \sum_{i=1}^n (f(\phi_i,\theta_i;\delta,\beta_1,\beta_2,\Gamma) - l_i )^2 
= \sum_{i=1}^n (\delta 
+ \beta_1\phi_i+\beta_2\theta_i
+  \frac{1}{2} (\phi_i,\theta_i)\Gamma
(\phi_i,\theta_i)^T 
- l_i )^2 \qquad ,
\end{equation}
and requires the minimization conditions
\begin{equation}
  \label{eq:kdchidais0}
  \frac{\partial\chi^2}{\partial \delta} 
=  \frac{\partial\chi^2}{\partial \beta_i} 
=  \frac{\partial\chi^2}{\partial \Gamma_{ij}} 
   \stackrel{!}{=} 0 \qquad .
\end{equation}
This leads to a six-dimensional inhomogeneous system of
linear\footnote{These equations are linear in the parameters $\delta$,
$\beta_1$, $\beta_2$, and $\Gamma$, which are to be determined.}
equations:
\begin{eqnarray}
  &&\left(\begin{array}{llllll}
      \sum       1           \hphantom{\theta_i }             
      & \sum  \phi_i  \hphantom{\theta_i }                      
      & \sum  \hphantom{\theta_i }     \theta_i 
      &\frac{1}{2}\sum   \phi_i^2  \hphantom{\theta_i }     
      & \sum  \phi_i \theta_i 
      &\frac{1}{2}\sum  \hphantom{\theta_i }      \theta_i^2
      \\
      \sum  \phi_i    \hphantom{\theta_i }       
      & \sum  \phi_i^2  \hphantom{\theta_i }                      
      & \sum  \phi_i\theta_i 
      &\frac{1}{2} \sum  \phi_i^3  \hphantom{\theta_i }     
      & \sum  \phi_i^2 \theta_i 
      &\frac{1}{2} \sum  \phi_i \theta_i^2
      \\
      \sum \hphantom{\theta_i }     \theta_i      
      &  \sum \phi_i\theta_i      
      & \sum  \hphantom{\theta_i }     \theta_i^2               
      &\frac{1}{2} \sum  \phi_i^2  \theta_i  
      & \sum  \phi_i \theta_i^2 
      &\frac{1}{2} \sum \hphantom{\theta_i }      \theta_i^3
      \\
      \sum  \phi_i^2      
      &  \sum \phi_i^3                   
      & \sum  \phi_i^2\theta_i 
      &\frac{1}{2}\sum   \phi_i^4  
      & \sum  \phi_i^3 \theta_i 
      &\frac{1}{2}\sum   \phi_i^2 \theta_i^2
      \\
      \sum \phi_i\theta_i 
      & \sum \phi_i^2\theta_i 
      & \sum \phi_i\theta_i^2
      &\frac{1}{2}\sum  \phi_i^3\theta_i 
      &\sum  \phi_i^2\theta_i^2 
      &\frac{1}{2}\sum  \phi_i\theta_i^3
      \\
      \sum  \hphantom{\theta_i }     \theta_i^2      
      & \sum  \phi_i\theta_i^2      
      & \sum  \hphantom{\theta_i }     \theta_i^3               
      &\frac{1}{2} \sum  \phi_i^2  \theta_i^2  
      & \sum  \phi_i \theta_i^3 
      &\frac{1}{2} \sum  \hphantom{\theta_i }     \theta_i^4
      \\
  \end{array}
  \right)
  \cdot
  \left(
    \begin{array}{c}
      \delta \\
      \beta_1 \\
      \beta_2 \\
      (\Gamma)_{11} \\
      (\Gamma)_{12} \\
      (\Gamma)_{22} 
    \end{array}
  \right)
 = 
  \left(
    \begin{array}{l}
     \sum   l_i      \\
     \sum   l_i \phi_i  \\
     \sum   l_i \theta_i  \\
     \sum   l_i \phi_i^2   \\
     \sum   l_i \phi_i \theta_i   \\
      \sum  l_i \theta_i^2   
    \end{array}
  \right)\nonumber \\
  \label{eq:solveme6}
\end{eqnarray}
One can choose the set of points $(\phi_i, \theta_i)$ in a suitable way to
simplify equation~(\ref{eq:solveme6}). If for each point $(\phi_i, \theta_i)$ 
also $(\phi_i, -\theta_i)$, $(-\phi_i, \theta_i)$, and $(-\phi_i,
-\theta_i)$ are added to the set\footnote{This is both a sensible and easily
fulfilled demand, as $\mathrm{h}_{\mathrm{best}}'$ corresponds to $\theta=\phi=0$.},
one arrives at
\begin{eqnarray}
  &&\left(\begin{array}{llllll}
      \sum       1           \hphantom{\theta_i }             
      & \hphantom{\phi_i} 0  \hphantom{\theta_i }                      
      & \hphantom{\theta_i} 0  \hphantom{\theta_i } 
      &\frac{1}{2}\sum   \phi_i^2  \hphantom{\theta_i }     
      & \hphantom{\phi_i} 0  \hphantom{\theta_i }
      &\frac{1}{2}\sum  \hphantom{\theta_i }      \theta_i^2
      \\
      \hphantom{\phi_i} 0 
      & \sum  \phi_i^2  \hphantom{\theta_i }                      
      & \hphantom{\phi_i} 0
      &\hphantom{\phi_i} 0
      & \hphantom{\phi_i} 0
      &\hphantom{\phi_i} 0
      \\
      \hphantom{\phi_i} 0
      &  \hphantom{\phi_i} 0
      & \sum  \hphantom{\theta_i }     \theta_i^2               
      &\hphantom{\phi_i} 0
      & \hphantom{\phi_i} 0
      &\hphantom{\phi_i} 0
      \\
      \sum  \phi_i^2      
      &  \hphantom{\phi_i} 0
      & \hphantom{\phi_i} 0
      &\frac{1}{2}\sum   \phi_i^4  
      & \hphantom{\phi_i} 0
      &\frac{1}{2}\sum   \phi_i^2 \theta_i^2
      \\
      \hphantom{\phi_i} 0
      & \hphantom{\phi_i} 0
      & \hphantom{\phi_i} 0
      &\hphantom{\phi_i} 0
      &\sum  \phi_i^2\theta_i^2 
      &\hphantom{\phi_i} 0
      \\
      \sum  \hphantom{\theta_i }     \theta_i^2      
      & \hphantom{\phi_i} 0
      & \hphantom{\phi_i} 0
      &\frac{1}{2} \sum  \phi_i^2  \theta_i^2  
      & \hphantom{\phi_i} 0
      &\frac{1}{2} \sum  \hphantom{\theta_i }     \theta_i^4
      \\
  \end{array}
  \right)
  \cdot
  \left(
    \begin{array}{c}
      \delta \\
      \beta_1 \\
      \beta_2 \\
      (\Gamma)_{11} \\
      (\Gamma)_{12} \\
      (\Gamma)_{22} 
    \end{array}
  \right)
 = 
  \left(
    \begin{array}{l}
     \sum   l_i      \\
     \sum   l_i \phi_i  \\
     \sum   l_i \theta_i  \\
     \sum   l_i \phi_i^2   \\
     \sum   l_i \phi_i \theta_i   \\
      \sum  l_i \theta_i^2   
    \end{array}
  \right) . \nonumber \\
  \label{eq:solveme3}
\end{eqnarray}
This leads to three equations:
\begin{eqnarray}
\label{eq:easy1}
\sum  \phi_i^2  \cdot \beta_1  &=& \sum   l_i \phi_i  \\
\sum  \theta_i^2  \cdot \beta_2  &=& \sum   l_i \theta_i  \\
      \sum  \phi_i^2\theta_i^2 \cdot (\Gamma)_{12} 
     &=& \sum   l_i \phi_i \theta_i   
\label{eq:easy3}
\end{eqnarray}
and a three-dimensional system
\begin{eqnarray}
  \label{eq:dreidimsys}
  \left(\begin{array}{llllll}
      \sum       1           \hphantom{\theta_i }             
      &\frac{1}{2}\sum   \phi_i^2  \hphantom{\theta_i }     
      &\frac{1}{2}\sum  \hphantom{\theta_i }      \theta_i^2
      \\
      \sum  \phi_i^2      
      &\frac{1}{2}\sum   \phi_i^4  
      &\frac{1}{2}\sum   \phi_i^2 \theta_i^2
      \\
      \sum  \hphantom{\theta_i }     \theta_i^2      
      &\frac{1}{2} \sum  \phi_i^2  \theta_i^2  
      &\frac{1}{2} \sum  \hphantom{\theta_i }     \theta_i^4
      \\
  \end{array}
  \right)
  \cdot
  \left(
    \begin{array}{c}
      \delta \\
      (\Gamma)_{11} \\
      (\Gamma)_{22} 
    \end{array}
  \right)
 &=& 
  \left(
    \begin{array}{c}
     \sum   l_i      \\
     \sum   l_i \phi_i^2   \\
      \sum  l_i \theta_i^2   
    \end{array}
  \right) \qquad ,
\end{eqnarray}
which can easily be solved.

The error estimators sought are contained in the parameters of
the paraboloid. The three degrees of freedom of $\delta$ and $\beta_{1,2}$
correspond to the position of the minimum of the curve. The three
parameters in $\Gamma$ contain the curvature of the paraboloid, which are
functions of the resolution, as the covariance matrix $C$ of the
Gaussian approximation is the inverse of $\Gamma$. From $C=\Gamma^{-1}$
one obtains:
\begin{eqnarray}
\sigma_{\phi}^2 &=& (\Gamma^{-1})_{11} \\
\sigma_{\theta}^2 &=& (\Gamma^{-1})_{22} \\
\mathrm{cov}(\phi,\theta) &=& (\Gamma^{-1})_{12}  \qquad .
\end{eqnarray}

\begin{figure}[tbp]
\begin{center}
\includegraphics[width=8cm]{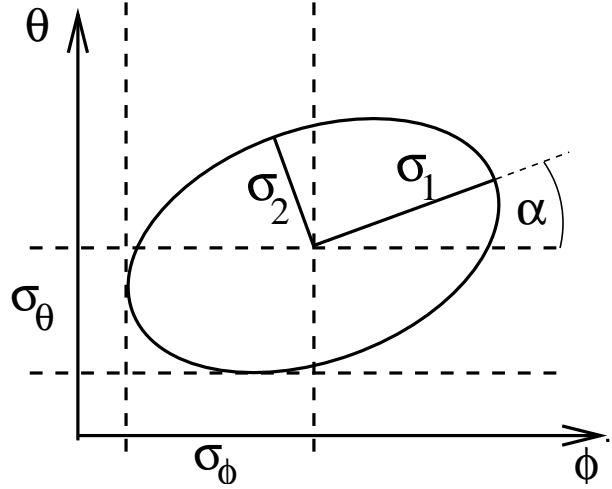}
\caption{The confidence ellipse in $\theta$ and $\phi$ can be represented
either by $\sigma_\theta$, $\sigma_\phi$, and the covariance
$\mathrm{cov}(\theta,\phi)$ or by the major and minor axes $\sigma_1$
and $\sigma_2$, and the rotation angle $\alpha$ of the ellipse between the
major axis and the $\theta$-direction.}
\label{fig:ellipsedefinitions}
\end{center}
\end{figure}
The results can also be represented in a more intuitive way using the
major and the minor axes $\sigma_1$ and $\sigma_2$ of the confidence
ellipse, and its rotation angle $\alpha$ with respect to the $\phi$-axis
(Figure~\ref{fig:ellipsedefinitions}). In order to find this alternative
representation, the covariance matrix $C$ needs to be diagonalized. That
leads to
\begin{equation}
\sigma_{1,2}^2 = \frac{\sigma_{\phi}^2 + \sigma_{\theta}^2}{2} \pm
\sqrt{\frac{(\sigma_{\phi}^2 + \sigma_{\theta}^2)^2}{4} + \det C} \qquad .
\end{equation}
Choosing $\sigma_1\ge\sigma_2$, the rotation angle $\alpha$ is then given
by the expression
\begin{equation}
\tan\alpha = \frac{\sigma_1^2 - \sigma_\phi^2}{\mathrm{cov}(\phi,\theta)}
\qquad.
\end{equation}
In the following discussion a third representation $(\sigma_{\mathrm{a}},
\epsilon,\alpha)$ will be used, where $\sigma_{\mathrm{a}}$ is connected to
the area of the ellipse and $\epsilon$ is its excentricity:
\begin{eqnarray}
\sigma_{\mathrm{a}} &=& \sqrt{\sigma_1\cdot\sigma_2} \\
\epsilon &=& \frac{\sigma_1}{\sigma_2} \qquad .
\end{eqnarray}
The rotation angle $\alpha$ is the same as in the previous parameterization.

This procedure can lead to unphysical, negative values for
the variance estimators. This happens whenever the basic track finding
procedure has not arrived at a good minimum in likelihood space. The
search for the minimum should then be repeated more carefully.

Note, that the resolution estimates are calculated for each pattern of hits
seperately. Still they are not the {\em true\/} values for the discrepancies
of true and reconstructed direction, but estimates of the {\em average\/}
discrepancy for a set of events, which are similar to the one under
consideration.

\section{Space angle resolution}
\label{section:spaceangle}

For a neutrino telescope it is common to quote the median $\bar{\Psi}$
of the spatial angle $\Psi$ between true and reconstructed track
direction in order to describe its resolution. This section is devoted to 
the connection of the error ellipse parameters to the space
angle uncertainty $\psi$. It is convenient to choose the representation of
the ellipse using $\sigma_{\mathrm{a}}$ and $\epsilon$, with $\alpha$ not
playing a role in this special context.

The true direction lies within the area given by the error ellipse
with a probability of $39.4\%$. The median of a distribution corresponds to a
$50\%$-probability. Hence as a first step the ellipses need to be
enlarged correspondingly.

In order to obtain $\psi$, one considers the Gaussian likelihood
function in its diagonal form, where the coordinates\footnote{These
coordinates $x_1$ and $x_2$ are connected to $\phi$ and $\theta$ by a
rotation of the angle $\alpha$.} are $x_1$ and $x_2$:
\begin{equation}
f(x_1,x_2) = \frac{1}{2\pi\sigma_1\sigma_2}
e^{-\frac{1}{2}\left(\left(\frac{x_1}{\sigma_1}\right)^2+\left(\frac{x_2}{\sigma_2}\right)^2\right)} 
\qquad .
\end{equation}
To obtain the median $\mathrm{M}$, the function $f(x_1,x_2)$ must
be integrated from the center $(0,0)$ outwards until a cumulated
probability of $0.5$ is reached. This integral is done in polar
coordinates (radius $\rho$ and angle $\gamma$)
\begin{equation}
\label{eq:aquivradius}
0.5 \stackrel{!}{=} \int_0^{\mathrm{M}}\int_0^{2\pi}
f\left(\rho\cos\gamma,\rho\sin\gamma\right)
\rho\,\mathrm{d}\gamma\,\mathrm{d}\rho  \qquad .
\end{equation}
For the special case of $\sigma_{\mathrm{a}} =\sigma_1=\sigma_2$ (i.e.
$\epsilon = 1$), the calculation can be carried out analytically and
yields $\mathrm{M} = 1.177\cdot\sigma_{\mathrm{a}}$.
\begin{figure}[tbp]
\begin{center}
\includegraphics[width=8cm]{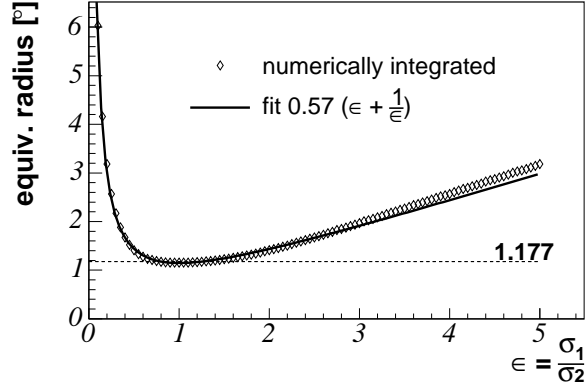} 
\caption{Obtaining the space angle resolution estimate by applying a
correction based on the excentricity. For fixed $\sigma_{\mathrm{a}} =
1^\circ$ and varying $\epsilon$, the {\em equivalent radius\/} $\mathrm{M}$ is
displayed, such that the true direction lies within a circle of radius
$\mathrm{M}$ with a probability of $50\%$.}
\label{fig:excen_corr}
\end{center}
\end{figure}

For the general case (with arbitrary $\epsilon$) the integration must be carried
out numerically. This has been done for the case of $\sigma_{\mathrm{a}} =
1^\circ$ and the result is shown in
Figure~\ref{fig:excen_corr}. 
The median $\mathrm{M}$ scales linearly with $\sigma_{\mathrm{a}}$. 

The curve is symmetric with respect to the exchange of $\sigma_1$
and $\sigma_2$. To improve the speed of the program an
approximation proportional to $(\epsilon + \frac{1}{\epsilon})$ has been
applied which results in the {\em excentricity corrected spatial angle
resolution estimation\/}:
\begin{equation}
\sigma_{a}^{\epsilon} = 0.57 \cdot (\epsilon + \frac{1}{\epsilon})\cdot
\sigma_a \qquad .
\end{equation}
The fit is also shown in Figure~\ref{fig:excen_corr}. The deviation in the
interval $\epsilon \in [1;3]$ is below $5\%$. For applications where this
is too coarse an approximation, the exact results of the numerical
integration can be stored in appropriate tables.

\section{Discussion}
\label{section:tests}

The above algorithms have been tested within the framework of the AMANDA
neutrino telescope. The
estimates perform in a stable and reliable way and produce sensible
results.

To ensure the correctness of the estimations, several testing procedures
both in data and in Monte Carlo have been carried out:

\begin{enumerate}
\item For $\sigma_{\theta}$ and $\sigma_{\phi}$ its {\em pull\/} has
been studied. It is defined as the ratio of the difference of true and
reconstructed direction over the resolution estimator. If the estimation
is good, the pull should be Gaussian distributed, centered at zero and
with unit width. This investigation can only be done in a Monte Carlo simulation.
\item For the spatial angle resolution the pull is not a sensible quantity,
because angles in space can only be positive. Hence one can study ensembles
of events with the same $\sigma_{\mathrm{a}}$ or
$\sigma_{\mathrm{a}}^{\epsilon}$, respectively. The median of the true
deviation can then be compared to the corresponding $\sigma_{\mathrm{a}}$.
Again this can only be done in the simulation.
\item In data the true directions are not known. Thus another approach is
being followed. Each event is split into two subevents by assigning every
second hit to subevent 1 and the remaining hits to subevent 2.
Each subevent undergoes reconstruction and is subjected to the
resolution estimation algorithm. That leads to directions\footnote{Note
that the reconstructed directions are expressed in standard detector coordinates,
whereas the error estimates are obtained in their respective rotated
systems. That is due to technical reasons only.} $\vartheta^i$,
$\varphi^i$ and errors $\sigma_{\theta}^i$ und $\sigma_{\phi}^i$ with
$i\in\{1,2\}$.

The pull
\begin{equation}
\mathrm{P}_{\vartheta} =
\frac{\mathrm{D}_{\vartheta}}{\sigma_{\theta}^{\mathrm{D}}}
= \frac
{\vartheta^1 - \vartheta^2}
{\sqrt{(\sigma_{\theta}^1)^2 + (\sigma_{\theta}^2)^2}}
\end{equation}
should also be a Gaussian distribution centered at zero with unit
width. For $\mathrm{P}_{\varphi}$ the difference in azimuth angles must
be multiplied by $\sin\vartheta$ to make up for the differences in the
rotated and not rotated coordinate systems:
\begin{equation}
\mathrm{P}_{\varphi} = \frac{\mathrm{D}_{\varphi}}{\sigma_{\phi}^{\mathrm{D}}}
= \frac
{(\varphi^1 - \varphi^2)\cdot\sin\vartheta}
{\sqrt{(\sigma_{\phi}^1)^2 + (\sigma_{\phi}^2)^2}}
\qquad .
\end{equation}
Of course this last check can also be done in the simulation.

\end{enumerate}

In all tests the results were satisfactory. 
Further details about the
tests and their results can be found in
\cite{phd:neunhoeffer_eng:2003}.

\subsection{Concluding remarks}

\begin{enumerate}
\item 
Often one applies quality cuts to the data set, which rely on additional
information that is not reflected in the construction of the likelihood
function itself. Hence this information is also not used in obtaining the
resolution parameters. In this case one expects the selected tracks to
be on average closer to the true direction than from the likelihood
analysis alone. Consequently one observes a pull with a width smaller
than 1.
\item 
The resolution parameters - such as $\sigma_{\mathrm{a}}$ - can
efficiently be used as cut variables, as misreconstructed events on
average have a worse resolution than correctly reconstructed ones.
\item 
The resolution estimation on an event-per-event basis can be used as
additional input in point source searches. The information is
hard to be included in the commonly used binned search algorithms
\cite{tillpub:ahrens:2003xy3}. An appropriate procedure based on maximum
likelihood methods that can integrate the new information in a natural
way will be presented in a forthcoming publication.
\end{enumerate}

\begin{ack}

I wish to thank the AMANDA collaboration for their support in obtaining
the method described herein. I thank Lutz K{\"o}pke and Alexander
Holfter for many an illuminative and productive discussion. I would also like
to thank the German Research Foundation (DFG) and the German Ministry
of Research and Education (BMBF) for financial support of the AMANDA
project.
\end{ack}

\bibliographystyle{plain}

\begin{thebibliography}{00}

\bibitem{art:recopaper:2004}
J.~Ahrens et~al.
\newblock {Muon track reconstruction in AMANDA}.
\newblock {\em Nucl. Instrum. Methods}, 2004.
\newblock in press.

\bibitem{tillpub:ahrens:2003xy3}
J.~Ahrens et~al.
\newblock {Search for extraterrestrial point sources of neutrinos with
  AMANDA-II}.
\newblock {\em Phys. Rev. Lett.}, 92:071102, 2004, astro-ph/0309585.

\bibitem{tillpub:Andres:2001ty}
E.~Andres et~al.
\newblock {Observation of high-energy neutrinos using Cerenkov detectors
  embedded deep in Antarctic ice}.
\newblock {\em Nature}, 410:441--443, 2001.

\bibitem{phd:neunhoeffer_eng:2003}
Till Neunh{\"o}ffer.
\newblock {\em {Development of a new method to search for cosmic neutrino point
  sources with the AMANDA neutrino telescope}}.
\newblock {Ph.D. thesis}, University of Mainz, Germany, 2004.
\newblock (Shaker Verlag, Aachen, ISBN 3-8322-2474-2, in German).

\end{thebibliography}

\end{document}